\documentclass[preprint]{aastex}
\pdfoutput=1
\usepackage{graphicx}
\usepackage{float}
\usepackage{color}
\usepackage[normalem]{ulem}
\usepackage{natbib} % included in AAS

\def\arcsec{\mbox{$^{\prime \prime}$}}

\def\deg{$^\circ$}

\def\nlfff{non-linear force-free field (NLFFF)}

\def\sdo{Solar Dynamics Observatory}
\def\hmi{Helioseismic and Magnetic Imager}
\def\aia{Atmospheric Imaging Assembly}

\usepackage{color}           % For color text: \color command

\shorttitle{Topological Study of AR 11158}
\shortauthors{Zhao et al.}

\begin{document}

\title{Temporal Evolution of the Magnetic Topology of the NOAA Active Region 11158}

\author{Jie {Zhao}\altaffilmark{1,2}, Hui {Li}\altaffilmark{1,2}, Etienne {Pariat}\altaffilmark{3}, Brigitte {Schmieder}\altaffilmark{3}, Yang Guo\altaffilmark{4}, Thomas Wiegelmann\altaffilmark{5}}
\email{nj.lihui@pmo.ac.cn}
\altaffiltext{1}{Purple Mountain Observatory, CAS, 2 West Beijing Road, Nanjing 210008, China}
\altaffiltext{2}{Key Laboratory of Dark Matter and Space Astronomy, CAS, Nanjing 210008, China}
\altaffiltext{3}{LESIA, Observatoire de Paris, Section de Meudon, F-92195,Meudon Principal Cedex, France}
\altaffiltext{4}{School of Astronomy \&\ Space Science, Nanjing University, Nanjing 210093, China}
\altaffiltext{5}{Max-Planck-Institut f\"{u}r Sonnensystemforschung, Justus-von-Liebig-Weg 3, 37077 G\"{o}ttingen, Germany}

\begin{abstract}

We studied the temporal evolution of the magnetic topology of the active region (AR) 11158 based on the reconstructed three-dimensional magnetic fields in the corona. The \nlfff\ extrapolation method was applied to the 12 minutes cadence data obtained with the \hmi\ (HMI) onboard the \sdo\ (SDO) during five days. By calculating the squashing degree factor Q in the volume, the derived quasi-separatrix layers (QSLs) show that this AR has an overall topology, resulting from a magnetic quadrupole, including an hyperbolic flux tube (HFT) configuration which is relatively stable at the time scale of the flare ($\sim 1-2$ hours). A strong QSL, which corresponds to some highly sheared arcades that might be related to the formation of a flux rope, is prominent just before the M6.6 and X2.2 flares, respectively. These facts indicate the close relationship between the strong QSL and the high flare productivity of AR 11158. In addition, with a close inspection of the topology, we found a small-scale HFT which has an inverse tear-drop structure above the aforementioned QSL before the X2.2 flare. It indicates the existence of magnetic flux rope at this place. Even though a global configuration (HFT) is recognized in this AR, it turns out that the large-scale HFT only plays a secondary role during the eruption. In final, we dismiss a trigger based on the breakout model and highlight the central role of the flux rope in the related eruption.

\end{abstract}


\section{Introduction}
It has been widely accepted that the different behaviors of the solar atmosphere in response to the plasma motions and instabilities are intensely related to the different topologies of the coronal magnetic field \citep{Berger1991}.Null points, separatrices and separators, where the connectivity of the magnetic field line is discontinuous, are classic topological structures that are preferential for the formation of electric current and the occurrence of energetic event. Flares and coronal mass ejections (CMEs) are the most energetic events taking place in the solar atmosphere and the energy needed to power them originates from the magnetic field in the corona. Some flares happen with null points in their associated magnetic configurations \citep[e.g.,][]{Aulanier2000,Demoulin2000,Manoharan2005,Jiang2013b} while others without any coronal null points \citep[e.g.,][]{Li2006b,Mandrini2006,Schmieder2007b,Chandra2011}. The concept of null point as well as separatrix has been extended to quasi-separatrix layer (QSL) in the last two  decades \citep[see review by][and references therein]{Demoulin2006}. QSLs are defined as 3D magnetic volumes with very sharp gradients of magnetic field connectivity \citep[e.g.,][]{Demoulin1996a,Titov2002}

QSLs are preferential sites for the build-up of electric currents and the development of magnetic reconnection \citep[e.g.,][]{Galsgaard2003,Aulanier2005,Pariat2006,Masson2009,Guo2013}. Narrow current sheets form dynamically in non-potential magnetic field and are responsible for flaring activity \citep{Aulanier2005}. Magnetic quadripolar configuration can enclose an hyperbolic flux tube (HFT) \citep{Titov2002}, which is a particular subdomain of the QSL and located at the intersection of several QSL branches. The HFT generalize the concept of magnetic separator. QSLs, especially the HFTs, are central in the flaring processes of AR \citep{Titov2002}. Multiple types of boundary motions can induce the spontaneous formation of strong currents layers at QSLs \citep{Demoulin1996a} which can induce reconnection when the current sheet width reach the dissipative scales \citep{Aulanier2005}. Magnetic pinching due to large-scale shearing motion at the photospheric footpoints of an HFT also causes the effective growth of current density in the HFT \citep{Titov2003}. Unlike reconnection happening at a 2D null point, where field lines reconnect in pair and change their connectivities suddenly, 3D reconnection at a QSL induce a continuous exchange of connectivity \citep{Aulanier2006}. Neighbouring field lines continuously exchange their connectivities within the QSL current sheet which induce an apparent slipping motion of the field lines relatively to each other \citep[see also][]{Aulanier2007,Torok2009,Masson2009,Masson2012}.

As reconnection occurs in the QSLs, particles accelerated by reconnection are flowing within the QSLs and therefore flare ribbons are expected to be cospatial with the photospheric footprints of the QSLs \citep[e.g.,][]{Demoulin1997}. Many observational studies have analyzed the intersection of the QSLs and the photosphere, and compared them to flare ribbons, providing indirect evidence of magnetic reconnections as the triggering mechanism of solar eruptive events \citep[][]{Schmieder1997,Mandrini1997,Mandrini2006,Masson2009,Chandra2011}

CMEs may severely affect the space environment and have been extensively studied both in theories and observations \citep[see review by][]{Forbes2006}. Reviews about the CME trigger processes can be found in \citet{Aulanier2013}. Three kinds of CME models have been generally accepted so far: the \lq\lq magnetic breakout\rq\rq  model \citep{Antiochos1999,Chenpf2000}, the \lq\lq tether-cutting\rq\rq model \citep{Moore1980,Moore2001,Lynch2004} and the \lq\lq flux rope\rq\rq models \citep[]{Forbes1991,Amari2000,Amari2004}. All these models have received various observational confirmations \citep[e.g.,][]{Aulanier2000,Guo2010,Guoyang2012,Chengxin2013a}.
Each of these models requires a current carrying structure, an ensemble of field lines storing non-potential energy, e.g., a twisted magnetic flux rope, or a sheared magnetic arcades system.  While the current carrying structure is the only requirement for the \lq\lq flux rope\rq\rq models, the other models require other type of topological configurations and are therefore not completely equivalent. The \lq\lq tether-cutting\rq\rq model involves magnetic reconnection bellow the current carrying structure. Therefore a topological structure that would allow current build-up, such as separatrices or QSLs, shall be placed bellow the current carrying structure. On the other hand, the \lq\lq magnetic breakout\rq\rq model requires the development of reconnection above or on the side of the current structure in order to partially open the magnetic configuration for the flux rope to erupt. Hence it requires the existence of overall global topological structure located above the non-potential structure. While in the original study a 2D null point was invoked, such topological structure could also be a separator or an HFT.

Observationally, active region generally presents multiple topological structures that can be consistent with different models. Several studies have addressed the specific topologies associated to diverse eruption models \citep[e.g.,][]{Ugarte2007,Aulanier2000,Guo2010,Guoyang2012,Chengxin2013a}. All these works usually only focus on one particular time of the pre-flare topology.  However, null points, separators, QSLs and HFTs do not magically appear in active regions. They are the consequence of the construction of the active region induced by flux emergence and by reconfiguration with the surrounding coronal field. Several numerical simulations have focus on the formation of these structures \citep[][]{Torok2009, Archontis2005, Galsgaard2007, Moreno-Insertis2013}. However, they are all based on an idealized magnetic configuration, which is a simplification of the real condition and can hardly represent the realistic magnetic field. To reveal the real procedure in solar atmosphere, data-driven simulation based on the photospheric magnetograms from observations has been suggested in \citet{Fanyl2011} (also see in \citet{Cheung2012,Jiangchaowei2012,Jiang2013b}). In the present study, we use a static approach by reconstructing magnetic field in the corona from the observations in the photosphere at different times before the flare. This work is a first attempt to follow the evolution of the topology of an active region in time, by computing time series of QSLs map from successive magnetic field extrapolation.

Our target is the NOAA 11158 active region, which has been remarkable for being the source of multiple violent active events. It has been extensively studied in various aspects, such as the sunspot motions \citep{Jiangyunchun2012}, the magnetic non-potentiality increase \citep{Vem2012}, its formation, and flare-associated magnetic field changes (like during the M6.6 and X2.2 class flares, see in Figure 1) \citep[][]{Liuchang2012,Wangshuo2012,Petrie2012,Gosain2012}.
However, the topology evolution has not been really followed even though \citet[]{Dalmasse2013} showed the existence of different quasi-separatrices. We focus on the topological analysis of AR 11158 during 3 days before the X2.2 flare and one day after it in this work. It is the first attempt to use a set of static topology analysis to study the evolution of the magnetic topology of an active region, which benefits from the 12 minutes cadence of the HMI vector magnetograms. Observations and data reduction, including non-linear force-free field (NLFFF) extrapolation method, are presented in Section 2. In Section 3, we introduce the method used to calculate $Q$ values and our results. We give our summary and discuss the physical issues of QSLs in Section 4.

\section{Observations and Data Reduction}
The vector magnetic field data and the extreme ultraviolet (EUV) images were taken respectively by the \hmi\ \citep[HMI;][]{Scherrer2012,Schou2012} and the \aia\ \citep[AIA;][]{Lemen2012} onboard the \sdo\ \citep[SDO;][]{Pesnell2012} on 12 - 15 February 2011. SDO observes the full Sun continuously with high resolution and high quality, allowing us to trace the studied AR 11158 from its first appearance on the disk until its disappearance from the disk.

The analyzed vector-magnetic-field data were taken with a spatial resolution of about 1\arcsec\ and at a cadence of 12 min, and, were processed using the standard HMI science data processing pipeline tool. The recorded Stokes parameters (I, Q, U and V) from the FeI 6173 \AA\ line were deduced with the so-called Very Fast Inversion of the Stokes Vector (VFISV) algorithm \citep{Borrero2011}. After resolving the 180\deg\ ambiguity of the transverse field with the ¡®minimum energy¡¯ method \citep{Metcalf2006,Leka2009}, the images were remapped with a Lambert cylindrical equal area (CEA) projection. Please refer to \citet{Scherrer2012} and \citet{Schou2012} for more information about the HMI instrument and to \citet{Liuyang2012} for how the vector magnetic field data were obtained. Data with CEA projection are used to conduct the NLFFF extrapolation to explore the magnetic configuration and topology in the corona. This is a valid approximation when the AR is not far way from the disk center and the selected region of interest (ROI) is within a moderate size.

The extrapolation to reconstruct the 3D magnetic fields in the corona from photospheric magnetograms was done with the optimization-based NLFFF algorithm proposed by \citet{Wheatland2000} and coded by \citet{Wiegelmann2004} after introducing a weighting function. The reconstruction of the coronal field is achieved by minimizing the objective function defined as
\begin{equation}
L=\int_{V}w(x,y,z)[B^{-2}|(\nabla\times\bold{B})\times\bold{B}|^{2}+|\nabla\cdot\bold{B}|^{2}]d^{3}x
\end{equation}
where, $w(x,y,z)$ is the weighting function. With $w(x,y,z)=1$, this approach becomes similar to %E changes to the case of
\citet{Wheatland2000}. Generally, observed photospheric magnetic field data are noisy and not necessarily force-free \citep[e.g.,][]{Gary2001}. Therefore, this method was further developed to include a preprocessing procedure to remove most of the net force and torque from the data, making the boundary more consistent with the force-free assumption \citep{Wiegelmann2006} and agree with the Aly's criteria \citep{Aly1989}. Even though the preprocessing does not remove all the net force, it does improve the extrapolation results. Reader are referred to the above-mentioned papers for the details about the extrapolation and the preprocessing methods.

The selected ROI for AR 11158 for the extrapolation consists of 534$\times$402 pixels and it was reduced to 178$\times$134 pixels with a pixel size of $\sim 1.1$ Mm after a 3$\times$3 pixel binning to save computing time and disk space. We apply the extrapolation method to the binned data, which yields a coronal field data cube of 178$\times$134$\times$134 pixels, i.e., the extrapolated data cube covers a volume of $\sim$ 195$\times$147$\times$147 Mm$^3$. Figure \ref{fig02} shows an example of magnetic field lines plotted for the extrapolated coronal magnetic field data cube at 23:58 UT on 14 February 2011. Different colors indicate different magnetic connectivity. Figure \ref{fig02} shows the complex magnetic configuration of the AR. The comparison between the extrapolated field lines and the observed coronal loops is referred to \citet{Sun2012a} since the NLFFF extrapolation method is the same.

The EUV images from AIA are exploited for comparison with the  HMI data after coalignment. AIA onboard the SDO satellite provides nearly simultaneous high spatial ($\sim$1.2\arcsec) and temporal ($\sim$12s) resolution full-disk images of the corona and the transition region in seven EUV, two ultra-violet (UV) and one visible wavebands. Emission for each of the EUV and UV wavebands is dominated by an emission line formed at a specific temperature. The temperature coverage of AIA is from 5$\times$10$^3$ to 2$\times$10$^7$ K. Images from 304\AA\ are used in this paper, corresponding to peak response temperature of 5$\times$10$^4$ K.

\section{Magnetic Topological Analysis}

To find the most likely place where reconnection occurs, a quantitative and topological term, squashing degree factor $Q$, was introduced by \citet{Titov2002} to describe the degree of change in the magnetic field line connectivity. It is a field-line invariant and measures the distortion of elementary magnetic flux tubes. The 3D volumes with high values of $Q$ ($>>$2) imply large changes of connectivity and define quasi-separatrix layers (QSLs). HFT is a particular subdomain of QSLs, located at the intersection of two QSLs and corresponding to the highest values of Q.  As discussed in the introduction, QSL and HFT are preferential sites for thin current layers to build-up and eventually for magnetic reconnection to develop. The computation of Q allows to determine the topology of a magnetic domain and to identify the main structures where reconnection preferentially occurs.

\subsection{Computational Method of QSLs}

The Q values in the photosphere and the plane 2.2 Mm above the photosphere (Section 3.2.2) are estimated through the Jacobian Matrix (M) of the field line mapping from one footpoint to another. Q is defined by the square of the norm of M, divided by the absolute value of the determinant of M \citep{Titov2002}. Here we use a uniform coarse grid, which is sufficient for the calculation of large-scale QSLs. Unless stated otherwise, the reference boundary for the evaluation of Q is the bottom boundary of the domain, i.e. the photospheric surface. We do not estimate the Q value of points whose field line does not reach the bottom boundary. These points subsequently appear in white or black in Figures \ref{fig04}, \ref{fig05}, \ref{fig06}. The grid size of the extrapolation box is $\sim 1.1$ Mm and the pixel size used to calculate $Q$ is 1/5 of the grid of extrapolation, $\sim 0.22$ Mm. The integration method for the magnetic field line is based on the fourth-order Runge-Kutta routines with a precision of 10$^{-4}$ for the position of the field line footpoint. This method limits the maximum value of $Q$ to $\sim10^{8}$. In these horizontal planes, $Q$ is computed for 705$\times$375 points within a region of 155$\times$82 Mm$^2$ in Section 3.2.1 and 3.2.2. The same resolution for the calculation of $Q$ within a subarea is used in Section 3.4.1 and 3.4.3.

Since magnetic field is of 3D nature, it is essential to detect the preferential places for current formation in a volume. Computing $Q$ value at every point in the volume is time-consuming (unless the resolution is relatively low), an alternative way is to focus on various vertical cuts in the volume. We use equation (20) proposed in \citet{Pariat2012} to calculate the distribution of $Q$ ($Q$-maps) along different vertical planes intersecting the 3D domain. The reference boundary for the evaluation of $Q$ remains the bottom/photospheric boundary. For all the vertical cuts perpendicular to X- and Y-axis (except the one in Section 3.4.2), an uniform coarse grid is adopted. The pixel size is $\sim 0.22$ Mm as aforementioned and the integration method for the field line is the same. For those parallel to the Y-axis, 375$\times$410 points corresponding to 82$\times$90 Mm$^2$ were computed. For the cuts parallel to the X-axis, 705$\times$410 points corresponding to 155$\times$90 Mm$^2$, were computed.

For the vertical cuts in Section 3.4.2, a multi-step procedure described in \citet{Aulanier2005} is adopted. Namely, we first compute $Q$ on a coarse grid, then proceed to higher resolutions for places where $Q$ is high until it converges to an almost fixed value. The integration method relies on Numerical Algorithms Group (NAG) routines which precise up to $10^{-10}$ for the position of the field line footpoint. This computation allows us to determine the values of $Q$ up to $\sim 10^{12}$. The computation is done with 513x513 number of points, corresponding to 18$\times$18 Mm$^2$.

\subsection{Topological evolution of AR11158}

\subsubsection{A Snapshot of Q Map in the Photosphere}

In order to better understand the evolution of the topology in time, let us first examine the particular photospheric Q map prior to the X-class flare. We present here the main topological structures that will be discussed in the following sections of this paper.
The photospheric $Q$ map at 23:58 UT on 14 February 2011 is displayed in Figure \ref{fig03}. A prominent QSL (Q0) appears around the PIL between P1 and N2. This QSL corresponds to some highly sheared structures (the field lines lie below the field lines in red of Figure \ref{fig02} and are studied in more details in Section 3.4.1). The surrounding structures are relatively more potential (field lines in red and cyan in Figure \ref{fig02}). Some strong QSLs are also present in the center (and around) of the main magnetic polarities. For example, QSL D2 is crossing the central part of the positive magnetic polarities P1 \& P2. The QSL D1 appears clearly in the middle of the polarity N2 and is also intersecting the center of N1. At the photospheric level these QSLs could appear disjointed and correspond to two different structures but we will see, using cross sections, that they are indeed part of the same unique topological structure (Section 3.2.3). In between N1 \& N2, the QSL D1 passes through a region of weak magnetic field, where several low lying structure are present, hence D1 appears discontinuous in Figure \ref{fig03}.

All these strong QSLs divide the volume in several domains having different quasi-connectivity. The different field lines in Figure \ref{fig02} present such domains. For instant, the footpoints of the field lines in purple (connects P2 to N2) are located right of D2 in the polarity P2 and above D1 in the negative polarity while the footpoints of the field lines in cyan (connects P1 to N2) are located above D2 in the polarity P1 and below D1 in the polarity N2, etc. For simplicity, we will concern about the two main QSLs D1 and D2, which primarily divide the magnetic volume into four quasi-connectivity domains (QCDs), i.e.,\ field lines in light seagreen (from P1 to N2; QCD1), in seagreen (from P1 to N1; QCD2), in purple (from P2 to N2; QCD3) and in olive (from P2 to N1; QCD4) in Figure \ref{fig02}. The relative locations of these QCDs in the volume will be clearly shown in Section 3.2.3.

These QCDs represent the main global domains. However the observational data are obviously more rich and complex and there are several other smaller QCDs, such as the cases of field lines in green, dark blue, black and pink in Figure \ref{fig02}. All these structures are bounded by QSLs of secondary importance, i.e.,\ less intense, and/or corresponding to relatively small volumes compared to QCD1-4. They are the consequence of the complexity of the magnetic field distributions that goes beyond the four main polarities. The field lines in these domains are usually low lying, and appear as low lying structures in the vertical Q map that we computed in sections 3.3. They are usually weak currents-carried and relatively few energy is stored in these domains. An exception is the field lines in green which are likely to be involved with the C-class flare around 06:28 UT on 14 February 2011 (see \citet{Dalmasse2013} for more details).

Large $Q$ values also appear at the boundary of Figure 3, all these features appear in weak magnetic field region in which many low altitude/small-scale structures are present, with lots of bald patches, low-lying nulls, etc. All these naturally create QSLs with very large value of $Q$, but their vertical extension in the coronal domain remains very limited (as confirmed by our vertical cut computation in Section 3.2.3). Since they are located in localized regions where the magnetic field is weak, we do not expect intense electric current build up in these locations. Therefore, they likely do not play any important role during the intense active events of the AR.

\subsubsection{Photospheric Q maps}

The evolution of $Q$ values in the photosphere during 12 -15 February 2011 is presented in Figure \ref{fig04}. The reference plane for the computation of Q is taken at Z=2.2 Mm. This enables to focus on the more prominent large-scale features, discarding low-lying QSLs \citep[e.g., as in][]{Savcheva2012}. Panel (g) is at the same time as Figure \ref{fig03}, but with much smoother structures here, since the features below Z=2.2 Mm are excluded from the calculation in this panel. The QSL (marked by yellow arrow) located along the PIL between P1 and N2 corresponds to Q0 and the other QSLs (marked by black arrows) located inside N2 and P1 correspond to D1 and D2, respectively, as shown in Figure \ref{fig03}. In fact, the existence of these QSLs (Q0, D1, D2) can be observed already at 23:58 UT on February 13 (panel d): D1 \& D2 remain stable at the time scale of the flare ($\sim$1-2 hours) from that time up to February 14 at 23:58 UT (panel g), even just after the M6.6 and X2.2 flares, which has been confirmed by our high-temporal (12 minutes) photospheric Q maps. However, we can not see the QSL Q0 before Figure \ref{fig04}(d), it appears when the shearing motion between P1 and N2 becomes important (relatively to the rest of the AR). The two main QSLs (D1 and D2) eventually formed from the 4 distinct QSLs that present at 23:58 UT on February 11 (marked by black arrows in Figure \ref{fig03}(a) \& (b)), which are located inside four polarities of the main quadrupole, can be identified from Figure \ref{fig04} (a) -- (c). This evolution is cotemporal of the merging of polarities. It should be noted here that the QSLs are moving from the side of the polarities to the center, e.g.,\ QSLs in N1, P1, N2 \& P2 are initially next to the drawn isocontours (see in Figure \ref{fig04}(a)) and can be observed to move in the center of each polarities. This illustrates the evolution of a system which is originally based on 4 polarities aligned, hence 4 parallel QSLs, to a magnetic quadripolar system with 2 main QSLs. The two main QSLs (D1 and D2) decayed (Q value diminished by about two orders of magnitude) almost 12 hours after the X2.2 flare (Figure \ref{fig04} (h)) and D1 almost disappeared one day later (Figure \ref{fig04} (i)). QSL Q0 seems still existing along the PIL but is relatively faint (Figure \ref{fig04} (h) \& (i)).

To sum up, there are three critical topological structures obtained from the photospheric Q maps, e.g., QSLs Q0, D1 \&D2. From the temporal series, QSL Q0 is prominent in Figure \ref{fig04} (d) and (g) (marked by yellow arrow) which are at the times just before the M6.6 and X2.2 flares, respectively. In our case, this means that the non-potentiality of magnetic field along Q0 is much larger in Figure 4(d) and (g). It is considered to be related to the high activity of AR 11158. The Q value of Q0 decreased after the onset of the M6.6 flare, which is induced by the release of the free energy and the diminishment of the non-potentiality. The morphology of Q0 is also quite modified after the flare (cf. Section 3.4.1). We need to point out that Q0 evolves on the time scale of the flare while D1 and D2 evolve on a larger time scale ($>$ 12 hours). The stable existence of the two main QSLs (D1 and D2), which mainly describe the different connectivity between field lines in purple (P2-N2), olive (P2-N1), light seagreen (P1-N2) and seagreen (P1-N1) in Figure \ref{fig02}, seems to indicate that the global disposition of the QSLs D1 and D2 retain its configuration before and after the flares. They are relatively unaffected by the active events.

\subsubsection{Evolution of the Topology in Vertical Q Maps}

$Q$ maps shown in Sections 3.2.1 and 3.2.2 only reveal the magnetic topology at the photospheric level. Here, we use equation (20) proposed in \citet{Pariat2012} to calculate $Q$ in vertical cuts (also see in Sections 3.3 and 3.4.2). The computational domain is 155 Mm$\times$82 Mm$\times$90 Mm.

Figure \ref{fig05} presents the temporal evolution of the $Q$ values of a vertical cut in a fixed plane at X=66 Mm. The time cadence is the same as in Figure \ref{fig04}. The two large scale QSLs (D1 and D2) identified in previous section are clearly visible here. At several locations they cross each other, dividing the volume into multiple subdomains of quasi-connectivity. The intersection of these QSLs corresponds to a region of particularly large Q values, an HFT. At the beginning of the evolution (23:58 UT on 11 February), the HFT is relatively low (less than 20 Mm) and goes upwards in the next 24 hours (up to more than 40 Mm). From 11:58 UT on 13 February, its height decreases and stays relatively stable (around 40 Mm) in the following 36 hours. The same tendency is also demonstrated by the height distribution of 75\% free energy accumulation (white curve in the bottom panel). This agreement hints that the free energy is mainly accumulated in the central quasi-connectivity domain (QCD1, see the panel at 23:58 UT on 14 February). The relationship between the four QCDs, which have been mentioned in Section 3.2.1, and the HFT is shown in the vertical cut at 23:58 UT on 14 February. A close inspection of the panels after 11:58 UT on 13 February reveals a very small structure in QCD1 (marked by yellow arrow). It appears just above the photosphere, then becomes pronounced at y=40 Mm at 23:58 UT on 14 February. This small feature, which footprints correspond to Q0, will be further discussed in Section 3.4.2.

The quadripolar configuration formed by two dipoles has been considered as a basic model for solar prominence and flares. Theoretical work of magnetic topology above a quadrupole has been done via modeling the corresponding sunspots by point-like sources \citep[]{Baum1980,Gorbachev1988,Titov2002}. As a result, nontrivial topological structure exists when the sources are on the photosphere \citep{Baum1980} while no null point (neither separator) exists when the sources are under the photosphere \citep{Gorbachev1988}. The computation of Q distribution is done by \citet[]{Titov2002} for the latter condition, which is more realistic by presenting distributed sources on the photosphere. The HFT configuration obtained from the extrapolation in this work is similar to the ideal result of \citet{Titov2002} (cf. Figure 9 in their paper), i.e.,\ two QSLs crossing each other in the vertical cut. We still see a small structure which is merely above the photosphere (marked by yellow arrow in Figure \ref{fig05}) in the central domain (QCD1), which is not seen in Figure 9 of \citet{Titov2002}. While the previous models were static, we have presented here for the first time the temporal evolution and the build-up of the topological configuration. Our investigation shows that the HFT was present for more than 72 hours before the flare and was stable since 11:58 UT on 13 February, i.e. for about 36 hours during which two large flares as well as many small flares happened.

\subsection{Topology before the X2.2 Flare}

To explore the magnetic topology before the X2.2 flare, the Q maps at 23:58 UT on 14 February (2 hours before the X2.2 flare) of the vertical cuts at different locations perpendicular to the X- and Y-axis are presented in Figure \ref{fig06}. The different QCDs are annotated in the panels. The central large-scale HFT is present in most of the cuts perpendicular to the X-axis (upper three rows of Figure \ref{fig06}) except the one at x=33 Mm (a1) and the one at x=110 Mm (h1). The vertical cut at x=99 Mm (g1) shows a secondary HFT (marked by white arrows) in it. Besides the higher HFT of the quadrupole, the lower one is induced by the existence of small-scale multipolar magnetic polarities at the photosphere in this region. There is a small structure (marked by yellow arrow) in the cuts at x=55, 66, 77 Mm (panel c1, d1, e1, respectively), whose height is less than 5 Mm. It may be related to the X2.2 class flare and we study it in more detail in Section 3.4.2.

The vertical cuts at various locations perpendicular to the Y-axis are shown in the lower three rows. The two main QSLs D1 and D2 are apparent in these cuts. They have no intersection in the vertical cut at y=37 Mm (b2), while they cross a bit in that at y=48 Mm (c2). In the vertical cut at y=59 Mm (d2), besides D1 and D2, a third parabolic-like curve appeared below the height of 20 Mm. It is induced by the different connectivity of small-scale polarities independent of the main quadrupole. In the vertical cuts at y=26 (a2) and y=70 Mm (e2), only either D1 or D2 is visible. Note that there is also an interesting feature (marked by yellow arrow) in the vertical cut at y=37 Mm, which is below the curve 'D2' and has an oval shape. This small feature only appears when the cut is crossing and parallel to the axis of a small-scale HFT (the one in Section 3.4.2).

From the analysis above, we see that the main 3D topological structures for the observed quadrupole is two quasi-separatrix domes which intersecting with each other in the Y direction. The QSLs D1 and D2 in all the slices always show the same structures (two domes) that we see with different angles. This finding is also consistent with the theoretical result in \citet{Titov2002} (cf. Figure 8 in their paper). However, there are some other structures manifesting the complex magnetic connectivity in the present observational data, e.g.,\ the lower HFT in Figure \ref{fig06} (g1) and the lower QSL beside D1 and D2 in Figure \ref{fig06} (d2). From the vertical cut at y=48 Mm (c2), we infer that the higher dome is relatively narrow and the lower one is wide since the curves D1 and D2 intersect at lower part. This is different from the result in \citet{Titov2002}, in which the higher dome is wide and the lower one is narrow. In addition, we can not find real separatrices or separator in present configuration, similarly to \citet{Titov2002}.

\subsection{Topological evolution around the X2.2 Flare}
In the following, we discuss the formation of a flux rope before the flare as well as the relationship between QSLs and flare ribbons around the X2.2 class flare.

\subsubsection{QSLs and Sheared Arcades}
To investigate the magnetic connectivity around the flare region, we zoom into the region inside the dashed rectangle in Figure \ref{fig03} and display the details in Figure \ref{fig07}. The field lines in green and yellow are mainly rooted in the three strong quasi-separatrix regions which encircled by three strong QSLs, respectively(marked by the black solid arrows), with the left two regions on QSL D1 and the right (with an elliptic shape) on QSL Q0 . They are long and intertwined, which located right above the strong QSL Q0. The field lines change their directions drastically when connecting to D1, at location (x=18,y=15), where a null point is found in our extrapolation. Part of the strongly sheared arcades represented by the field lines in blue lies above the flux rope (see Figure \ref{fig07}).

The evolution of the topology around the flare time is presented in Figure \ref{fig08}. The aforementioned strong quasi-separatrix regions on D1 (left arrows) and Q0 (right arrow) change considerably after the flare (see that at 01:22 UT and 02:22 UT). For example, the elliptic region (marked by the black arrow on right) had already shrunk immediately after the flare. Therefore the structure to which it is related, the twisted flux rope, must have dynamically evolved during the flare. Here we suspect that the flux rope erupted and produced the flare, which will be further confirmed in Section 3.4.2.

\subsubsection{Flux rope topology}
The Q maps of the vertical cuts along white lines in Figure \ref{fig08} are shown in Figure \ref{fig09}. $Q$ values are calculated with a high spatial resolution on non-uniform grids as mentioned in Section 3.1. We see a small-scale HFT exists at 23:58 UT on 14 February, more than two hours before the X2.2 flare. It has an inverse tear-drop shape which gives the boundary of the highly sheared system and the surrounding arcades. There are also small features inside the inverse tear-drop structure, indicating the existence of different quasi-connectivities inside the sheared system (e.g.,\ twisted flux rope in yellow and green and sheared arcades in blue in Figure \ref{fig07}). These field lines are relatively low with an average height of 2 Mm and the intersection of QSLs (i.e.,\ HFT) is below 1 Mm. The inverse tear-drop structure disappeared after the flare, since the flux rope is not there, indicating the eruption of the flux rope. The QSLs remained at 02:22 UT likely correspond to only some sheared arcades with very low height.

We need to emphasize that the inverse tear-drop is a special shape of QSL when a flux rope is present. The fact that there is a closed QSL indicates a region with a different connectivity from all its surrounding, which is what a twisted flux rope gives. \citet{Pariat2012} investigated the QSLs of the magnetic configuration of \citet[][TD hereafter]{Titov1999}, which has a twisted flux rope with number of turn Nt=2. Their results (see in Figure 5 in their paper) show that an inverse tear-drop QSL wraps around the twisted flux rope and crosses itself in the bottom of the flux rope at an HFT. Similar result can also be found in \citet{Savcheva2012}, in which they obtained this special structure around a flux rope with Nt=4/3 also from TD model. In \citet{Guo2013b}, there are only bald patches (locations on the PIL where coronal field lines threading through tangentially) at the bottom of the flux rope, no closed QSL can be identified since the obtained flux rope touched the photosphere in bald patches and is not strongly twisted. Hence, not strongly twisted flux rope do not have closed inverse drop QSL. Figure \ref{fig09} undoubtedly prove that a strongly twisted flux rope appears in our case. The fact that this inverse tear-drop structure had disappeared after the flare is a clear topological indication that the flux rope had disappeared. In the 2 hours interval of the flare, the photospheric motions are far too limited for a drastic change of topology. We observe ribbons at the location of the QSLs, so the flux rope must be resistively activated as the erupting flux rope interacted with the above field lines by magnetic reconnection at the large-scale HFT. A CME was observed to originate from this particular AR in correlation with the flare (see Table 1 in \citet{Vem2012}), hence a twisted flux rope must have been ejected from this AR.

\subsubsection{QSLs and Flare Ribbons}

To reveal the relationship between the QSLs and the flare ribbons, AIA intensity images in 304 \AA\ is superimposed with the $Q$ maps around the X2.2 flare in Figure \ref{fig10}. The AIA image in the upper left panel is taken at 01:50 UT and the overlaid $Q$ value is at 01:22 UT on 15 February. The flare ribbons are located on both sides of Q0, mainly in between QSLs D1 and D2 in the central part of the map except the upper left part of the upper ribbon. \citet{Janvier2014} also studied this flare ribbons and compared this J-shaped ribbons with the standard 3D flare model. Their results show close correspondence between the observation and the model. Another ribbon (marked by black arrow in the lower left corner) presents along the QSL D1 with a crescent shape. The lower left panel shows the UV enhancement at 02:22 UT on 15 February, about half an hour after the flare peak time. The ribbons expanded after the flare onset, and they are mostly along QSLs D1 and D2 at this time. In more details, there is also brightening along the upper right part of D2, which is considered to be unrelated to the expansion of flare ribbons. If we check the Q maps, the QSLs D1 and D2 also separate a bit after the flare, as D1 moves towards north and D2 to south.

Based on the obtained results, we propose a possible scenario for the flux rope formation and eruption. The photospheric shearing motion between P1 and N2, which induced by the flux emergence, produced highly sheared arcade around the PIL. \citet{Liuchang2012,Liuchang2013} found evidence that the field lines of the sheared arcade underwent a tether-cutting mechanism for the M6.6 flare. Since the shearing motion existed during almost the whole lifetime of this AR, the similar mechanism would probably work for the X2.2 flare. From topological view, electric currents are preferentially forming in the small-scale HFT \citep{Aulanier2005} and tether-cutting reconnection is supposed to happen when the resistive instability (such as tearing instability, see in \citet{Barta2011}) occurs in the currents. This reconnection would braid the sheared arcades into a flux rope. Like the procedure that has been suggested in \citet{Aulanier2010}, the flux rope would rise slowly to a critical point where the torus instability causes the full eruption. In 2D the erupting flux rope would directly interact with the overlying HFT and reconnection would be going on there. In 3D things are much more complicated. In some cases, the erupting flux rope can simply slide next to the HFT. There is no need that the HFT is forced directly: the CME can pass next to it and the HFT would eventually be on the side, which has been partly shown in several simulations (e.g., \citet{Lugaz2011}).

The flare ribbons here are initially involving QSL Q0. They are supposed to be produced when the flux rope, which lying along Q0, erupted and reconnected with the overlying field lines (such as field lines in red in Figure 2). The chromospheric observation of the flare ribbons proves that QSLs D1 and D2 are not activated, i.e., no energy is flowing within them induced by reconnection at the large-scale HFT, around the trigger moment of the flare. They are only activated towards the end of the evolution of the flare, as is indicated with the co-spatial location of the brightening with the upper right part of D2. This may be associated with the interaction of the erupting flux rope with the large-scale HFT. The field lines inside the HFT (such as field lines in light seagreen, seagreen, purple and olive) will probably reconnect with each other when the flux rope keeps going upward and push the field lines above together. It is notable that this reconnection may happen just at the end of the flare, which manifests that the reconnection is not a trigger but a result of the eruption. For the breakout model, the reconnection above needs to happen before or/and during the eruption, which is not the same as in our condition. The right panels in Figure \ref{fig10} display the horizontal components of magnetic field and the corresponding $Q$ values at 01:22 UT and 02:22 UT, respectively. The horizontal field strength (marked by white arrows) near the PIL, which coincides with QSL Q0, increases after the flare, consistent with \citet{Fengli2013}. The location of the enhancement is also below the small-scale HFT (see in Section 3.4.2).

\section{Discussion and Conclusion}

In this paper, we conducted topological analysis of AR 11158 based on the computed $Q$ values. The findings are summarized and discussed as follows:

Based on the vertical cuts of $Q$ maps perpendicular to the X- and Y-axis, we describe the main topology of AR 11158 at 23:58 UT on 14 February as an HFT configuration, which is the typical topological structure of a quadrupole. This configuration is similar to the ideal result obtained by \citet{Titov2002}, i.e.,\ two quasi-separatrix domes crossing each other in the corona. Some other structures also appeared beside the main HFT configuration, as a result of the complex connectivity of observed magnetic field, which are not identified in the magnetic configuration of the modeled point-like sources.
From the temporal series of $Q$ maps in 3D, we noticed that the quadrupole is relatively stable and lasted for at least 36 hours, starting from 11:58 UT on 13 February, which was there even during the large flares. From the photospheric traces of the HFT configuration (Figure \ref{fig04}), we see that two main QSLs (D1 and D2) are stable since 11:58 UT on 13 February. An elongated QSL Q0 (marked by yellow arrows in Figure \ref{fig04} (d)), which is always located in the vicinity of the PIL, is considered to be related to the high activity of AR 11158. From the coronal traces of the HFT configuration (Figure \ref{fig05}), large-scale HFT always exists (before 11:58 UT on 15 February) and is stable for 36 hours since 11:58 UT on 13 February.

\citet{Titov2003} also studied this kind of large-scale HFT structure and suggested that the magnetic pinching inside caused by the large-scale shearing motion on the photosphere could produce a large flare. However, in our case, there is no large flare directly related to this HFT. The possible reconnection at this place may induce the UV enhancement at the intersection of HFT and the photosphere (i.e.,\ D1 and D2), which were not identified in the beginning of the flare in our observation. Hence, QSLs D1 and D2 are not activated during the X2.2 class flare. Some brightening appears at the end of the flare along QSL D2, indicating reconnection eventually happened at the large-scale HFT. It is possibly induced by the interaction between the erupting flux rope and the HFT. The evolution of the emerging flux region and the initiation of eruption has been investigated through MHD simulations with modeled magnetic configuration which is a simplification of real observation. With photospheric magnetograms, the data-driven MHD simulations may give a more realistic result. To the best of our knowledge, our work is probably the first investigation of the topological evolution of emerging flux region from observation.

We also investigated the QSLs related to the X2.2 class flare, which is the first X-class flare in solar cycle 24. From the $Q$ map in the photosphere at 23:58 UT on 14 February, a strong QSL (Q0) exists between P1 and N2, which corresponds to some highly sheared structures that include the twisted flux rope (i.e.,\ field lines in green and yellow in Figure \ref{fig07}) as well as the short sheared arcades (i.e.,\ field lines in blue in Figure \ref{fig07}). From the evolution of the photospheric traces of the magnetic topology (see in Figure \ref{fig08}), the shape of the QSLs which are located around the footpoints of the twisted flux rope changes drastically around the flare time, indicating that the eruption involves these field lines.

A small-scale HFT, which has an inverse tear-drop shape and is about 0.4 Mm above the photosphere, appears 2 hours or even longer before the flare. The inverse tear-drop structure disappears just after the flare peak, which confirms the eruption of the twisted flux rope. The remained QSLs after the flare probably correspond to sheared arcades with very low height.  \citet{Savcheva2012a} also found an HFT structure with a height of around 3 Mm under a long-lasting sigmoid, which exists for several hours before the eruption. Investigation of coronal flux rope formation mechanism and eruption through MHD simulations suggested the formation of an HFT before eruption \citep{Aulanier2010}.

The flare ribbons in the 304 \AA\ image at 01:50 UT on 15 February (six minutes after the flare onset) are located on both sides of QSL Q0 and between the two QSLs (D1 and D2). This manifests that the ribbons are initially involve QSL Q0 while D1 and D2 are not activated during the flare. According to previous studies \citep{Demoulin1997,Mandrini1997,Bagala2000,Masson2009}, the UV flare ribbons are usually located next to, or along the chromospheric traces of QSLs. The different locations referring to the closest QSLs may be caused by the separating motion of ribbons, as they expand after the flare onset \citep[e.g.,][]{Aulanier2012}. However, the ribbons will be confined inside the closest large-scale QSLs \citep{Chenpf2012,Guoyang2012}. In our observation, the flare ribbons stopped at the photospheric traces of the large-scale HFT (i.e.,\ D1 and D2).

In this study, we have obtained two HFT structures in our extrapolation, the large-scale one is in the corona while the small-scale one is in the chromosphere. There is no evidence for the strong current formation in the large-scale HFT before the X2.2 flare since no corresponding brightenings are found at the photospheric traces of this HFT before the flare. Besides, the large-scale HFT remains relatively stable while the small-scale flux rope associated HFT changes drastically around the flare. Hence, this large-scale HFT may only play a secondary role in the eruption. While our investigation would dismiss a trigger based on the breakout model, the precise identification of the trigger mechanism as well as the flux rope formation goes beyond the present study. Our topological analysis however clearly identifies a twisted flux rope priori to the flare which disappears after and which is related to the initial flare ribbons. The trigger is therefore likely to involve this particular twisted flux rope.

\acknowledgements

The data have been used by courtesy of NASA/SDO and the HMI science team. SDO is a mission of NASA's Living
With a Star program. We would like to thank Dr P. D{\'e}moulin for fruitful discussions and cordial supports. We benefit a lot by talking with other colleagues at Observatoire de Paris in Meudon.  We also thank Dr. L. Feng for reading and improving the manuscript as well as helpful discussion. J. Zhao and H. Li are supported by the National Basic Research Program of China under grant 2011CB811402, by NSFC under grant 11273065, and by the Strategic Pioneer Program on Space Sciences, Chinese Academy of Sciences, under grant XDA04071501. J. Zhao is also supported by NSFC under grant 11303100. Y. Guo is supported by the NSFC under the grant numbers 11203014, 10933003, and the grant from the 973 project 2011CB811402.

\bibliographystyle{apj}
%\bibliography{references} % file containing the bibtex ref (.bib)

\begin{figure}%[H] %fig01
\centering
\includegraphics[width=14.cm]{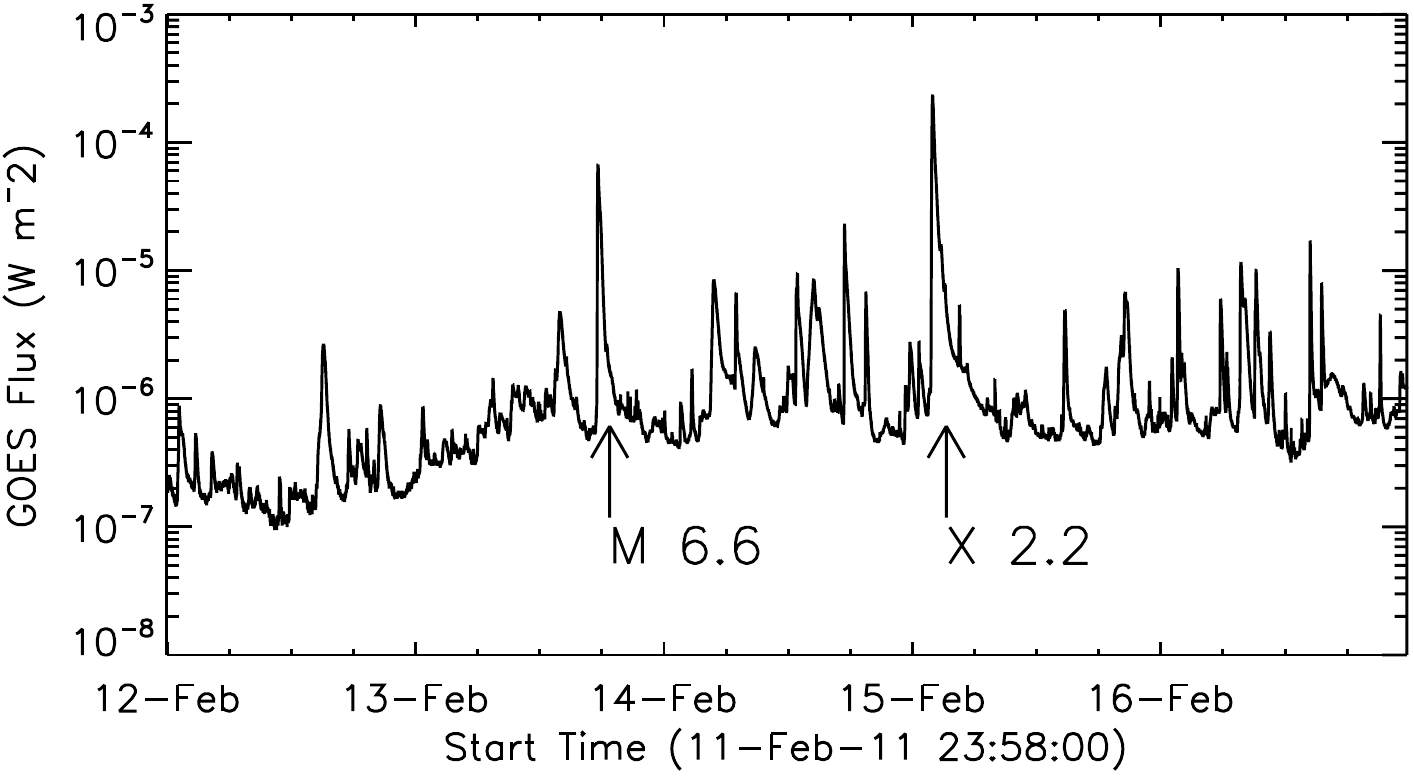}
\caption{GOES 1 -- 8 \AA\ X-ray flux for 5 days.}
\label{fig01}
\end{figure}

\begin{figure}%[H] %fig02
\centering
\includegraphics[width=8cm]{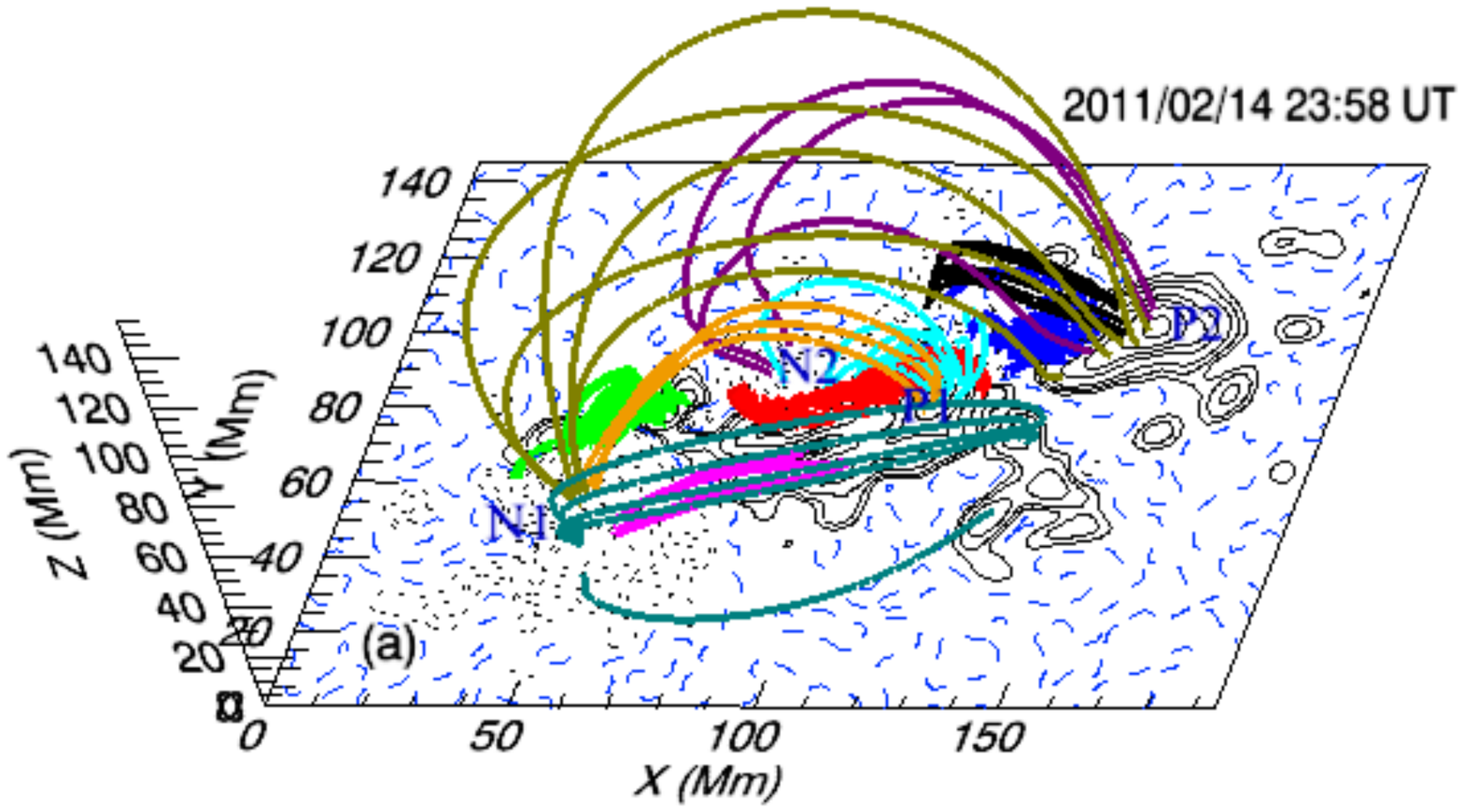}
\includegraphics[width=6cm]{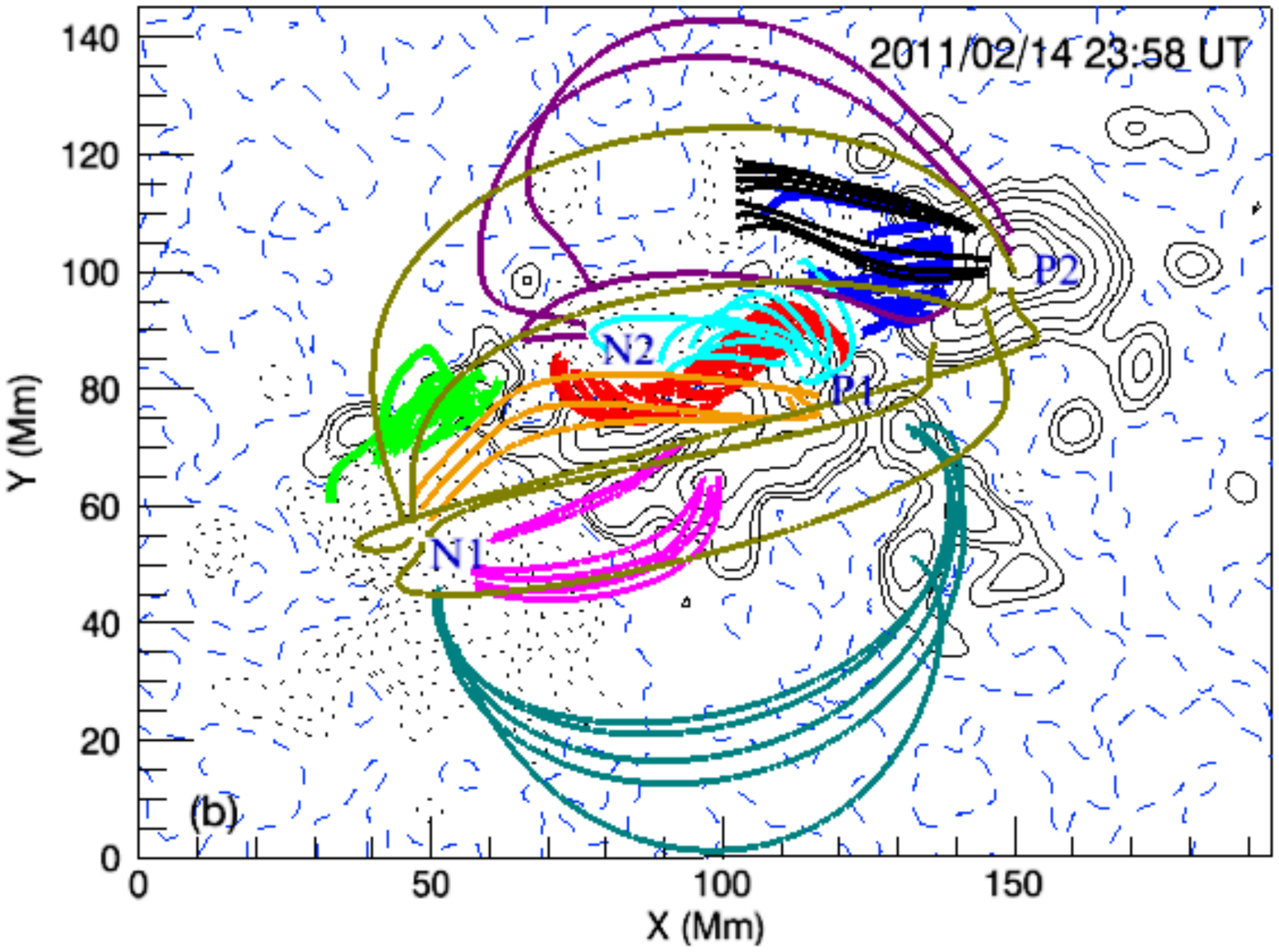}
\caption{Example of magnetic field lines of force drawn from the reconstructed coronal fields at 23:58 UT on 14 February 2011, which outlines the complex connectivity of AR 11158. (a) bird-view; (b) top-view. The solid and dotted contours show the positive and negative vertical magnetic field, respectively, with contour level of $B_z$ = $\pm$[50,100,200,500,800,1200,1800,2400,3000] Gauss. N1 and N2 mark negative polarities while P1 and P2 for positive ones.}
\label{fig02}
\end{figure}

\begin{figure}%[H] %fig03
\centering
\includegraphics[width=14.cm]{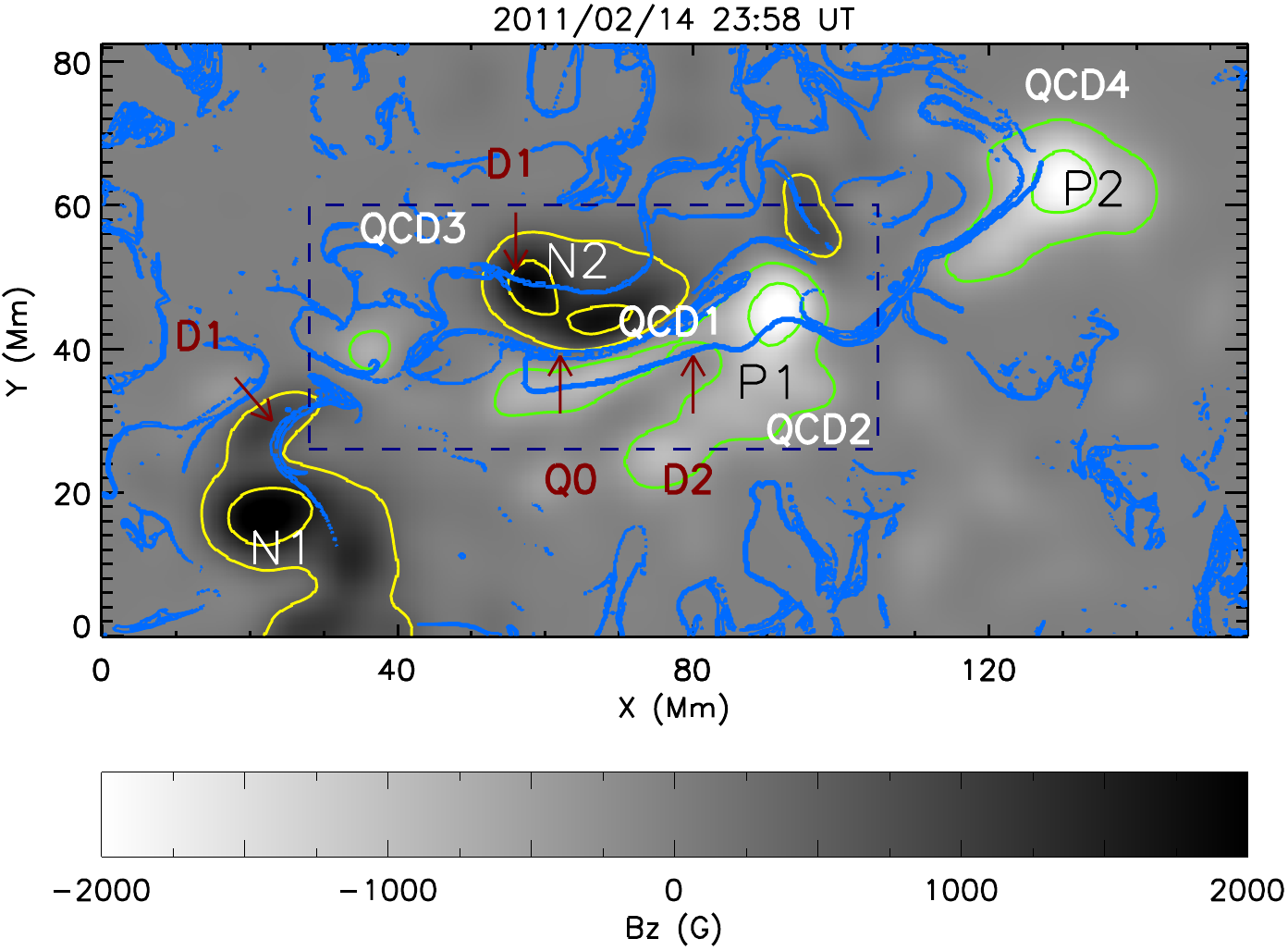}
\caption{Vertical magnetic field ($B_z$) in the photosphere at 23:58 UT on 14 February 2011, 2 hours before the peak time of the X2.2 class flare. The blue lines correspond to the isocontours of the squashing degree factor $Q$, indicating the locations of the main QSLs in the region. The green and yellow contours show the positive and negative vertical magnetic field, respectively, with contour level of $B_z$ = $\pm$[500, 1500] Gauss. N1 and N2 mark negative polarities while P1 and P2 for positive ones. Q0, D1, \& D2 mark the main QSLs while QCD1-4 locate the main quasi-connectivity domains (see description in Section 3.2.1).}
\label{fig03}
\end{figure}

\begin{figure}%[H] %fig04
\centering
\includegraphics[width=16.cm]{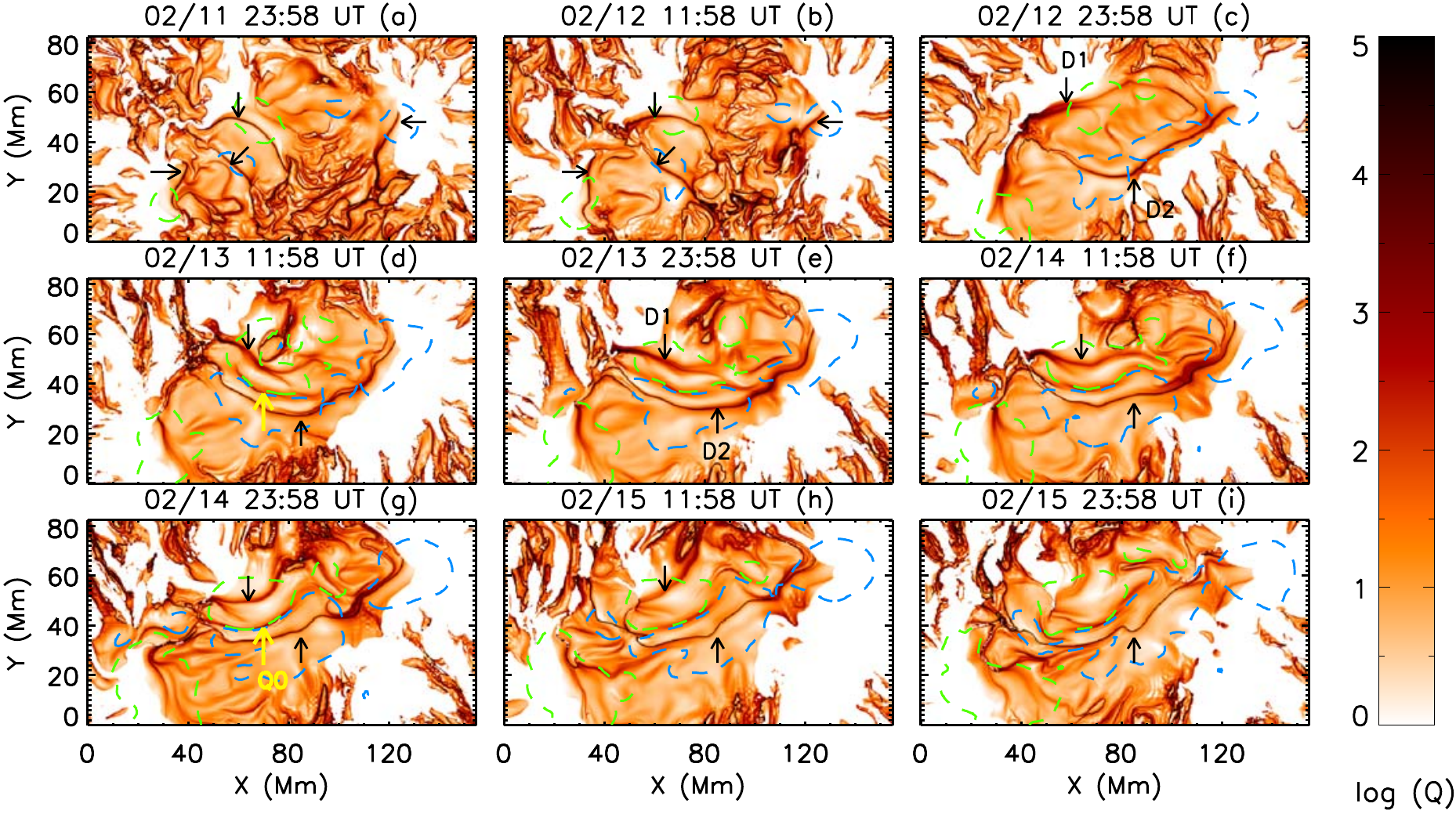}
\caption{Evolution of $Q$ maps at z=2.2 Mm during 4 days. The $Q$ values are displayed in logarithm scale. The FOV is the same as in Figure \ref{fig03} and time difference between each two successive panels is 12 hours. The blue and green contours show the positive and negative vertical magnetic field at z=2.2 Mm with level of $\pm$200 Gauss. The white regions in the map indicate places where field lines go out of the extrapolation box.}
\label{fig04}
\end{figure}

\begin{figure}%[H] %fig05
\centering
\includegraphics[width=12.cm]{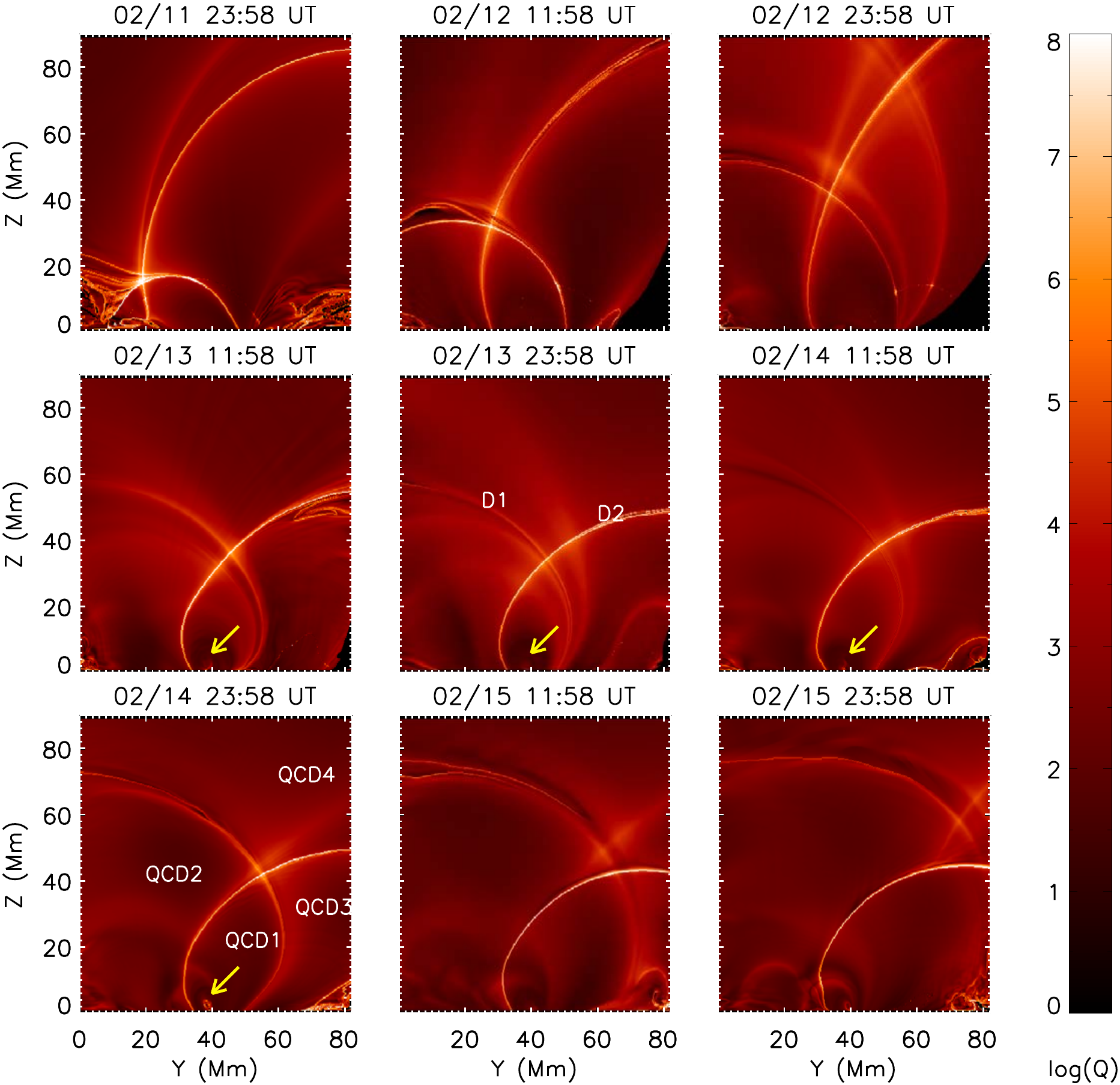}
\includegraphics[width=12.cm]{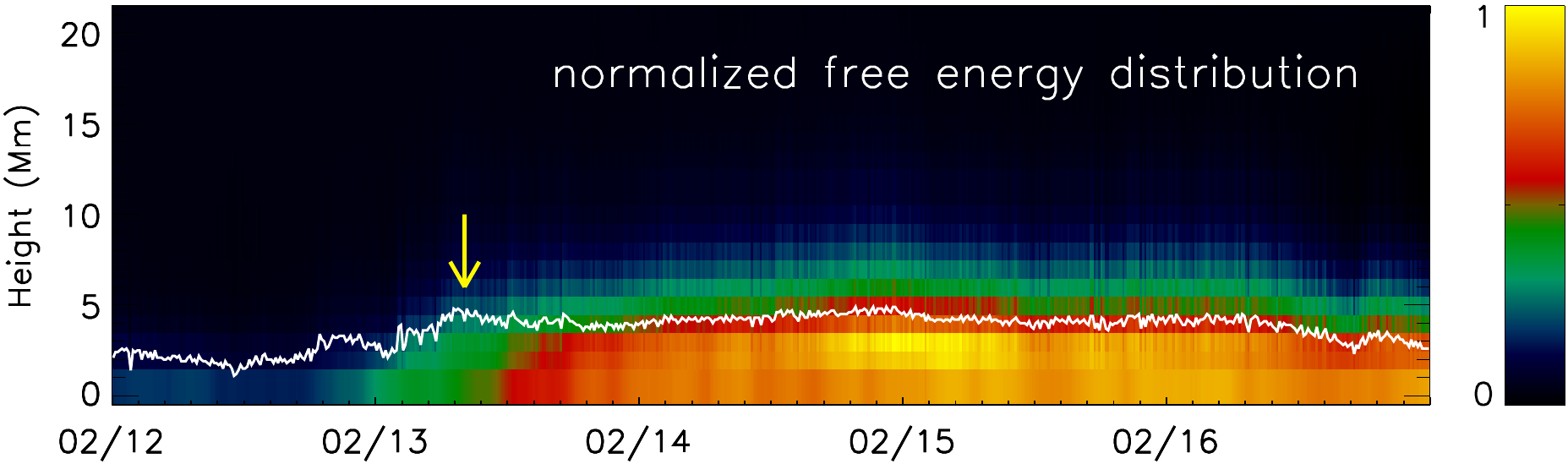}
\caption{Top: Evolution of $Q$ maps in a vertical cut at X=66 Mm. Time difference between two successive panels is 12 hours. Bottom: Temporal evolution of free energy distribution along height. The white curve indicates the height where 75\% free energy accumulated.}
\label{fig05}
\end{figure}

\begin{figure}%[H] %fig06
\centering
\includegraphics[width=11.cm]{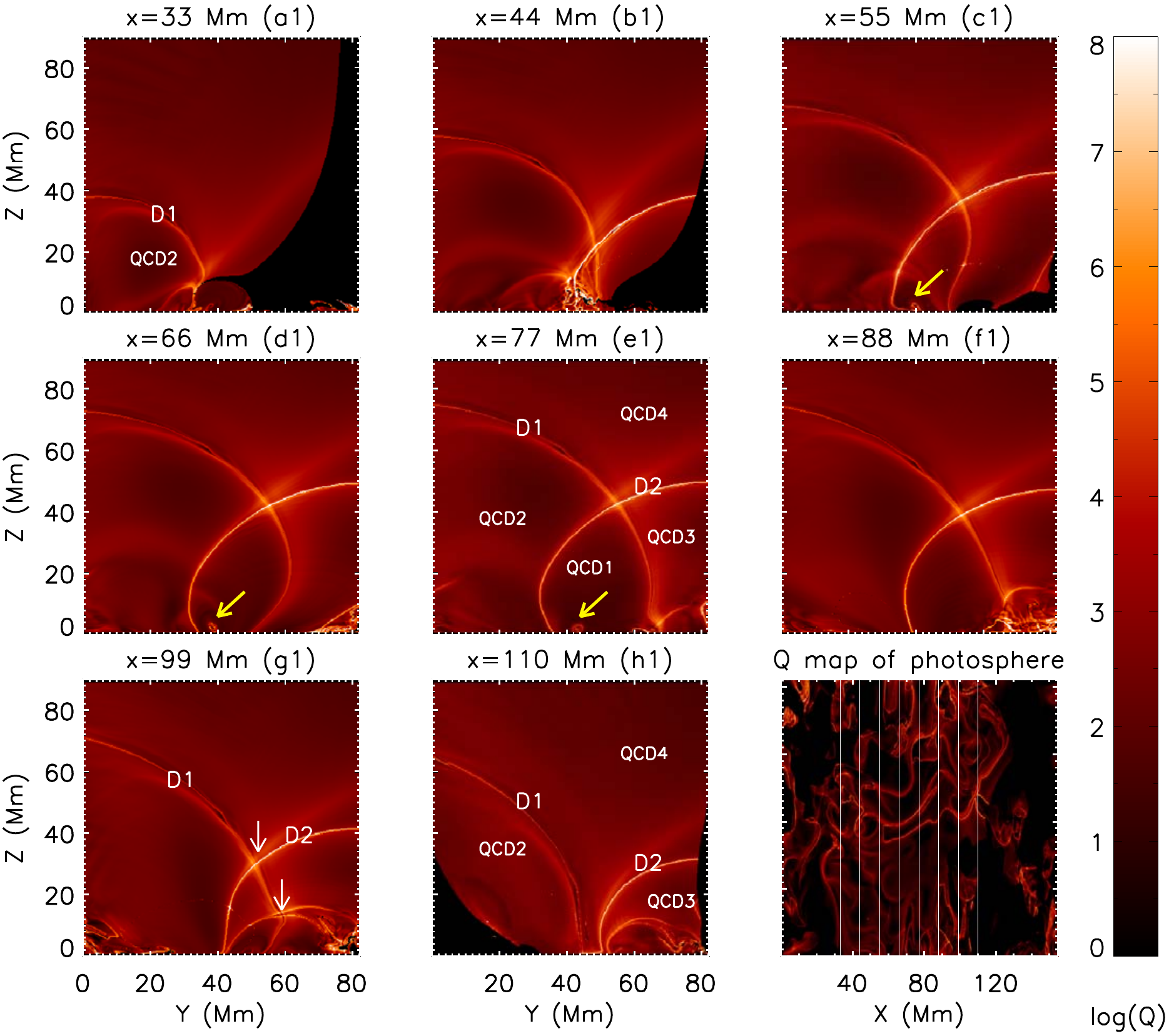}
\includegraphics[width=11.cm]{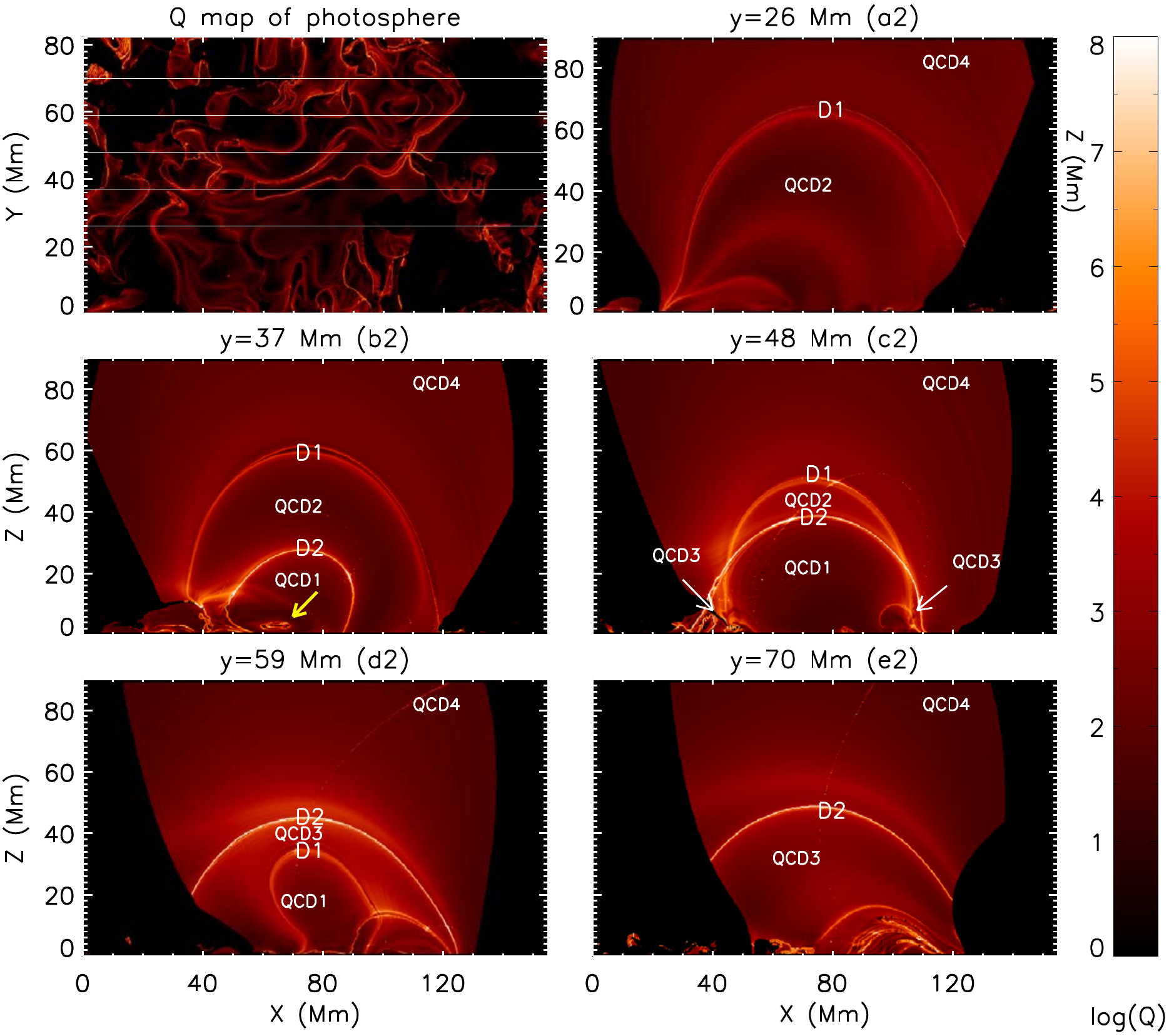}
\caption{$Q$ values on the plane of vertical cuts at different locations perpendicular to the X- (upper panels) and Y-axis (lower panels) at 23:58 UT on 14 February. The white vertical and horizontal lines indicate the locations of vertical cuts perpendicular to the X- and Y-axis, respectively.}
\label{fig06}
\end{figure}

\begin{figure}%[H] %fig07
\centering
\includegraphics[width=14cm]{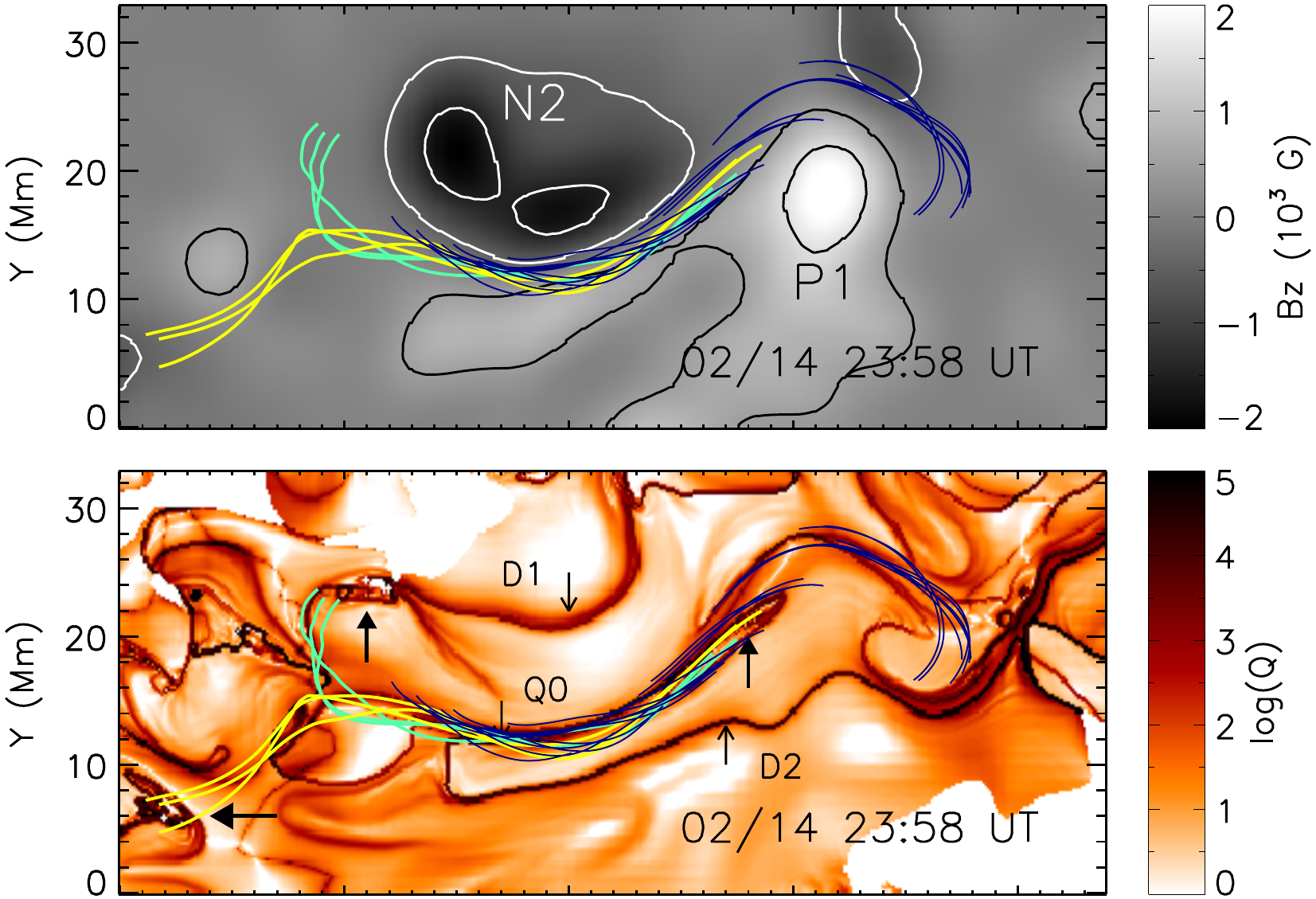}
\includegraphics[width=6.5cm]{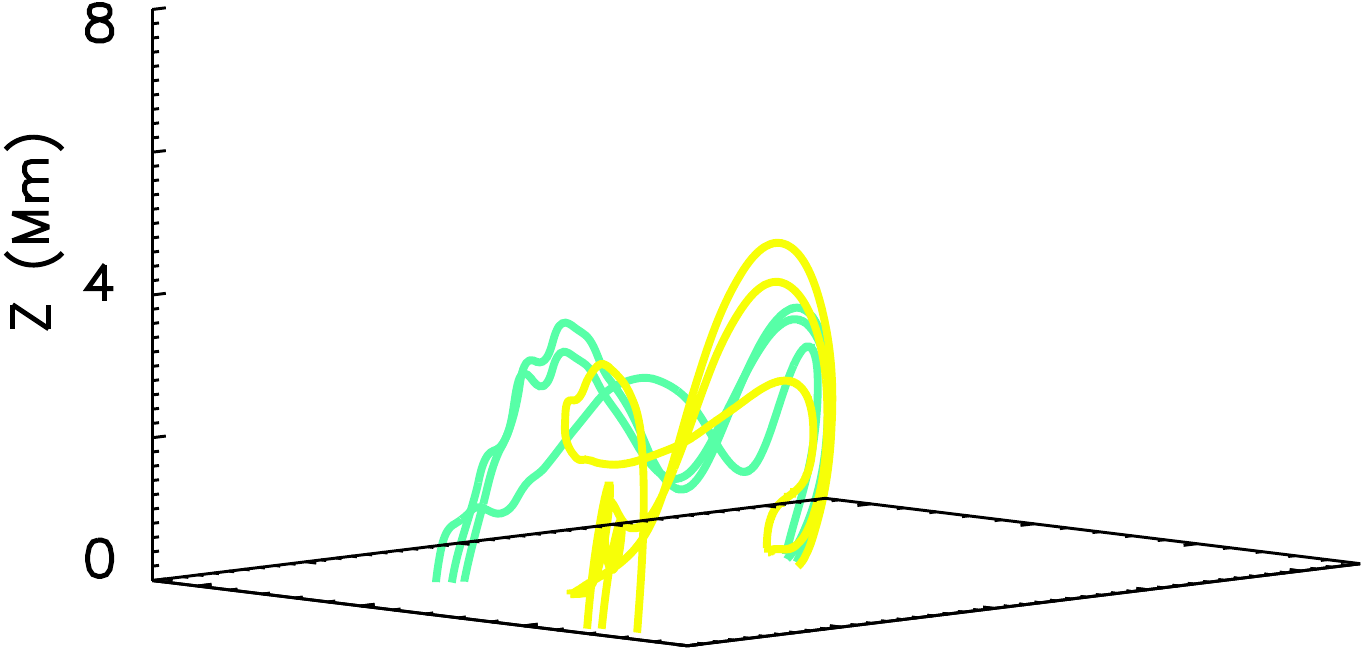}
\includegraphics[width=6.5cm]{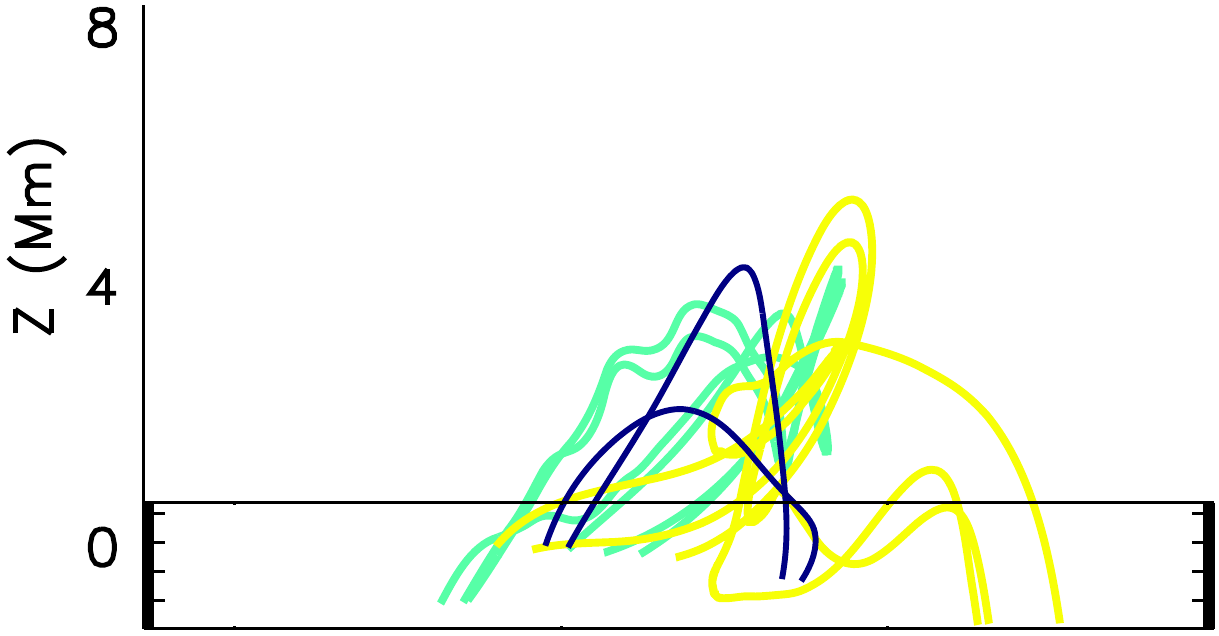}
\caption{Top: Bz map with overplotted magnetic field lines at 23:58 UT on 14 February. The FOV is delineated by the dashed rectangular box in Figure \ref{fig03}. Middle: $Q$ map with overplotted magnetic field lines. Bottom: Two perspective views of the field lines plotted in upper panels, with the same FOV in the X-Y plane.}
\label{fig07}
\end{figure}

\begin{figure}%[H] %fig08
\centering
\includegraphics[width=12.cm]{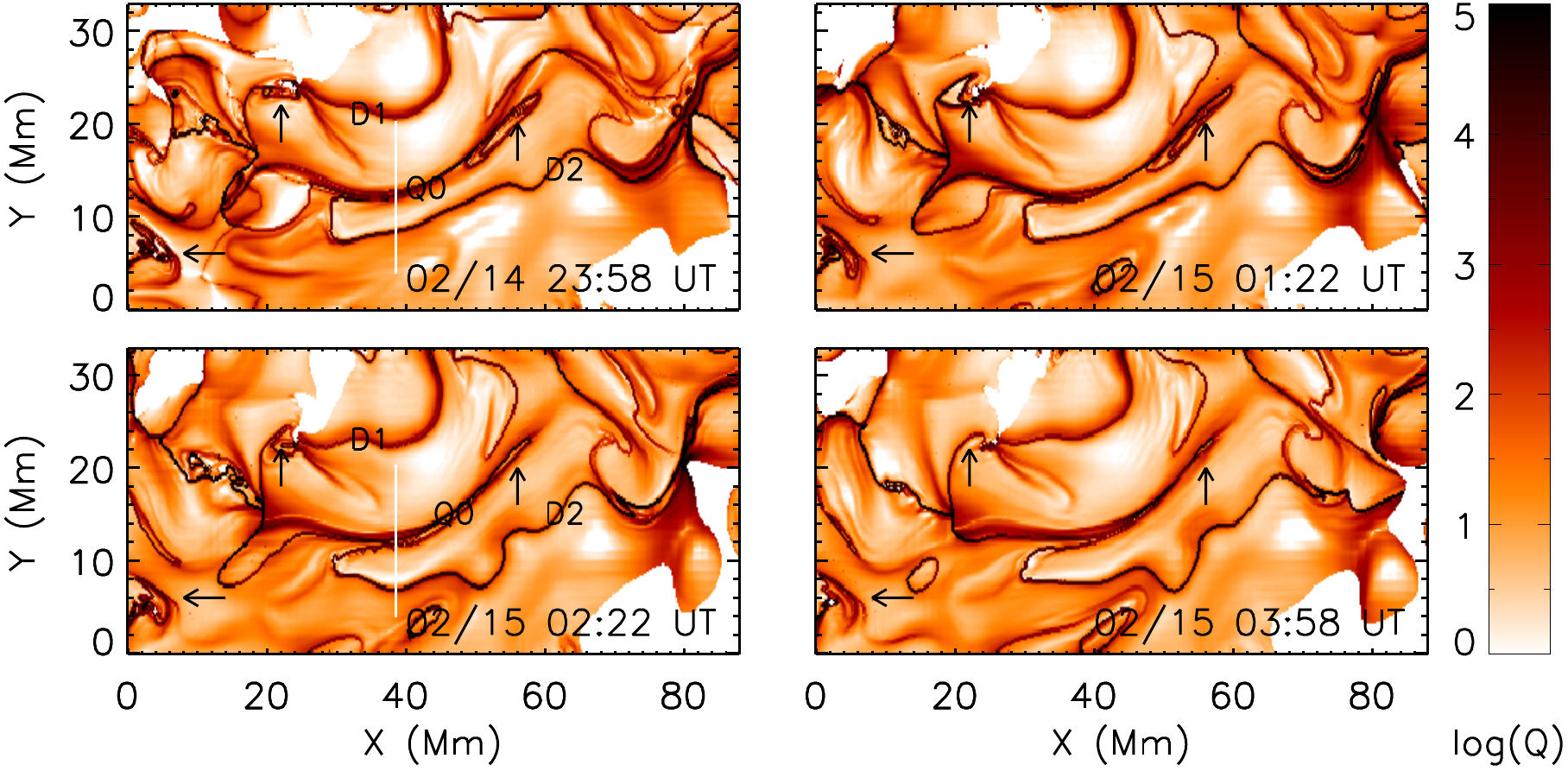}
\caption{The $Q$ values at four different times around the X2.2 flare. The FOV is the same as the upper panels in Figure \ref{fig07}.}
\label{fig08}
\end{figure}

\begin{figure}%[H] %fig09
\centering
\includegraphics[width=12.cm]{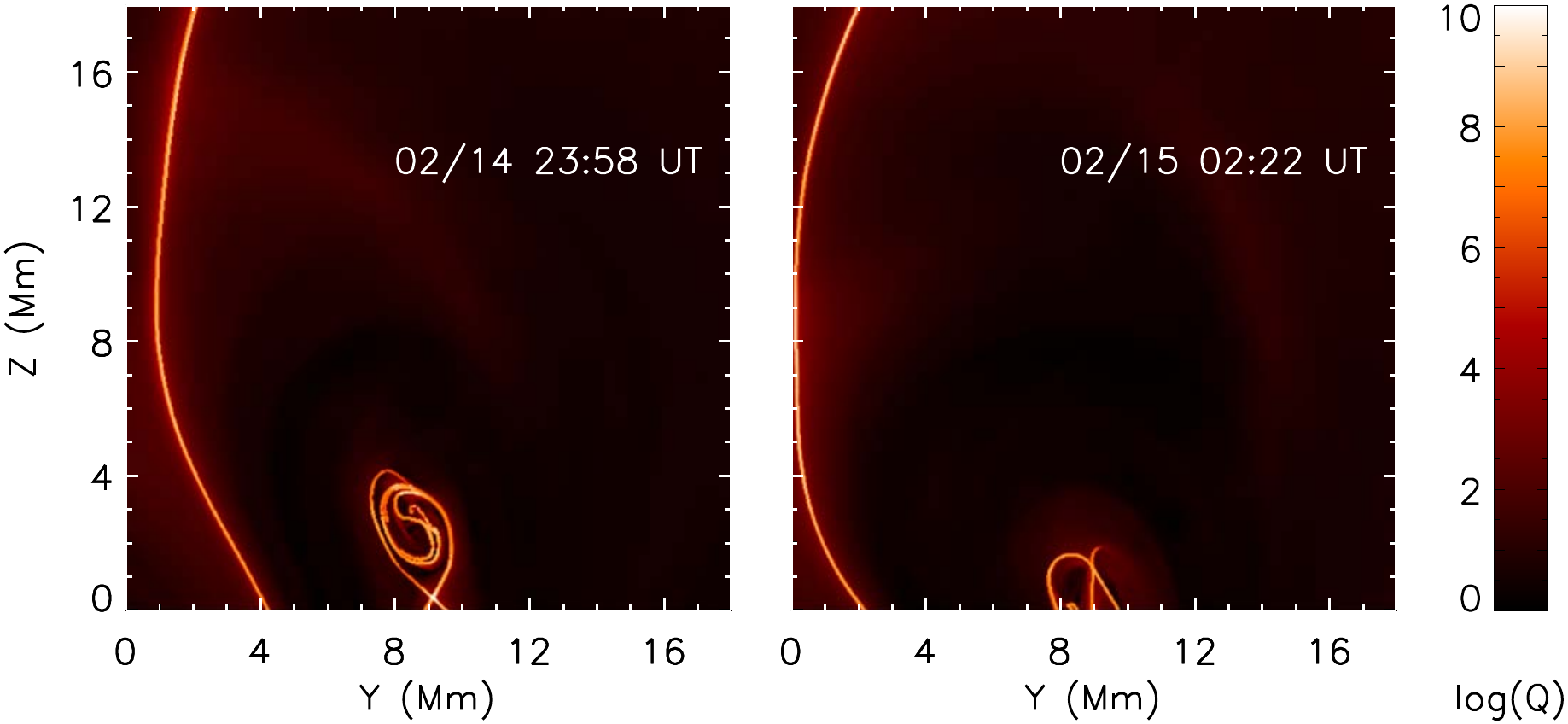}
\caption{$Q$ maps before and after the X2.2 class flare for the vertical cuts shown in Figure \ref{fig08} by the white lines.}
\label{fig09}
\end{figure}

\begin{figure}%[H] %fig10
\centering
\includegraphics[width=16.cm]{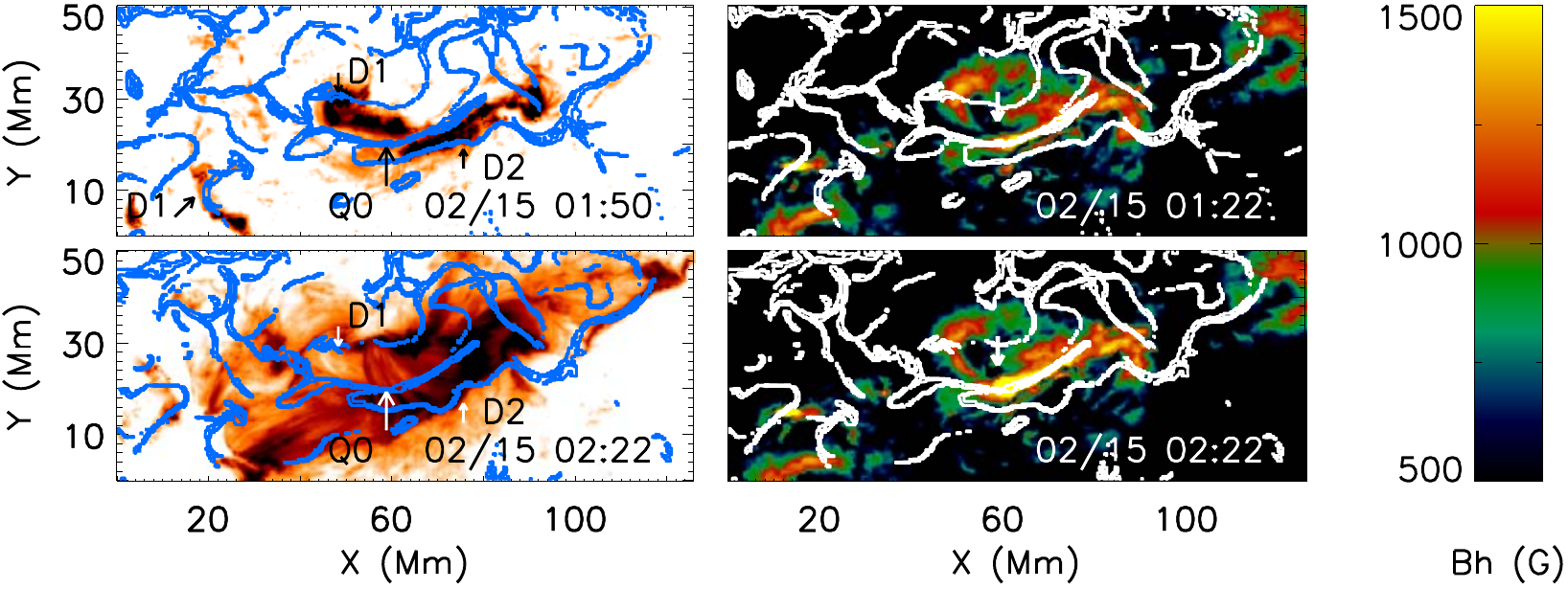}
\caption{Left: AIA 304\AA\ images at 01:50 UT (upper panel) and 02:22 UT on 15 February (lower panel), superimposed with $Q$ values (contour). Right: $Q$ values are superposed on the horizontal magnetic field at 01:22 UT (upper panel) and 02:22 UT on 15 February (lower panel). The blue and white contours in the panels indicate the main QSLs in this region. The Q maps in the upper/lower panels are at 01:22 UT/02:22 UT in the lower panels on 15 February, respectively.}
\label{fig10}
\end{figure}

\end{document}